%% file: paper.tex
\documentclass{aa}
\usepackage[dvips,xdvi]{graphics}

\def\Msun{$M_\odot$}
\newcommand{\Teff}{\mbox{$T_{\rm eff}$}}
\def\CH4bds{$CH_4BDs$}
\def\CH4bd{$CH_4BD$}
\newcommand{\J}{\mbox{\rm J}}

\begin{document}

  \thesaurus{06 % Formation structure and evolution of stars
	     ( 08.12.2;  % Stars: low-mass, brown dwarfs
	       08.12.3;  % Stars: luminosity function, mass function
               13.09.6)  % Infrared: stars
            }
   \title{The density of methane brown dwarfs: observational
 and theoretical constraints
	  }

   \author{
           F. D'Antona \inst{1}
	   E. Oliva \inst{2}, and
           A. Zeppieri \inst{1}
           }

\offprints{F. D'Antona %\\ email: dantona@coma.mporzio.astro.it
           }

   \institute{Osservatorio Astronomico di Roma,
	      Monte Porzio Catone, I-00040 Roma, Italy
\and
Osservatorio Astrofisico di Arcetri, Largo E. Fermi 5, I-50125 Firenze, Italy
              }

 \date{Received 18 June 1999; Accepted 1 October 1999}

   \titlerunning{The density of CH$_4$ Brown Dwarfs}

   \maketitle

\begin{abstract}
Methane brown dwarfs (\CH4bds), i.e. objects similar to the low mass star
Gliese 229B, are characterized by very unusual colours: J--K$\la$0 and
I--J$\ga$5. An analysis of the ESO public images, which cover an area of 37
sq. arcmin in the three filters,  yields one clear \CH4bd\ candidate with
J=20.2 and two fainter sources with J$\simeq$23. The resulting observed
density of methane dwarfs ranges from $\approx$100 to $\approx$500
objects per sq. degree with J$<$23.1, quite high but compatible, within
the errors, with what we derive from simulations of the stellar population of
brown dwarfs in the disk of the Galaxy adopting reasonable (although still
speculative) hypotheses on the evolution of BD colours with cooling. The
predictions presented here can be useful in constraining the results from
future searches of these objects in the infrared.
Deep imaging of several sq. degrees of sky down to J$\simeq$22,
K$\simeq$22 and I$\ga$27 are necessary to significantly improve the
observational database.
\end{abstract}

\section{Introduction}

Until recently, only three very low luminosity objects were known with
spectral type later than M9.5, but the two near IR ground based surveys DENIS
and 2MASS opened a new window in the study of the end of the main sequence
and its continuation into the realm of Brown Dwarfs (BDs). In fact, DENIS
added three objects (Delfosse et al. \cite{Delfosse}, 
Tinney et al. \cite{Tinney}), and 2MASS (Kirkpatrick et al. \cite{Kirkpatrick}) 
increased this sample till now by about 20 objects
spectroscopically confirmed by the Low Resolution Infrared Spectrometer
at Keck. These findings are defining a new spectral class, the ``L
dwarfs" spanning \Teff$\sim$2200--1500 K, their exact temperatures
depending on the treatment of dust in model atmospheres.

It was recently recognized that the near IR colours of the late M dwarfs
(\Teff$\la$2600 K) could not be reproduced by model atmospheres,
unless they included consideration of grains. Although the spectral feature
are very complex, the first generation of model atmospheres including grain
condensations 
(Tsuji et al. \cite{Tsuji}, Allard \cite{Allard}) 
are able to interpret the main
emerging spectral characteristics, and include predictions also for the
L--dwarf spectra.
Dust grains forming in the upper layers of very cool dwarfs have a large
opacity, and produce a redistribution of the IR flux (greenhouse effect),
explaining the very red J--K colours of the latest M and of the L-dwarfs.
The L--dwarfs clearly define the low end of the MS, which, depending on the
opacity, and thus on the dust inclusion in the models, may be shifted even
as low as $\sim$1800 K (Burrows and Sharp \cite{burrows-sharp}). 
However, due to the
steepness of the mass luminosity relation around the hydrogen burning minimum
mass (D'Antona \& Mazzitelli \cite{DM85}), 
the range of \Teff's of L-dwarfs
should be mostly populated by cooling brown dwarfs. An indication confirming
this interpretation is the presence of Lithium in about half of the 2MASS
objects, indicating that they are unable to burn their initial lithium, and
thus that their mass is below $\sim$0.06 $M_\odot$ (see however Allard 
\cite{Allard} for
a different interpretation).

When L-dwarfs evolve further, they will eventually reach
\Teff$<$1500 K. Just as for normal white dwarfs, the evolutionary time
is dictated by the release of the thermal energy of the ions, so that
it increases when decreasing the luminosity. Only one object  was until
recently identified as the result of such cooling: the companion to the low
mass star Gliese 229, having a spectrum showing the CH$_4$ bands, predicted
by theory. This object, the ``Rosetta stone" of cool BDs, 
at \Teff$\sim$900 K and luminosity well below $10^{-5}L_\odot$\ 
(Leggett et al. \cite{Leggett99}), 
shows in fact a peculiar spectrum characterized by metal depletions,
and dominated by H$_2$O, CH$_4$ and alkali metals. The flux
redistribution at very low \Teff\  produces a spectrum much bluer in
J--K, while keeping a very red I--K colour 
(e.g. Burrows et al. \cite{burrows97}). Therefore,
while the typical colour of L--dwarfs are J--K$>$1.3  
(e.g. Leggett et al.  \cite{Leggett98}), 
the cooler BDs should have J--K$\simeq$0, as confirmed by the colour
of Gliese 229B 
(Nakajima et al. \cite{nakaj}, Legget et al. \cite{Leggett99}).

Although this outline of low mass and brown dwarf evolution is qualitatively
sound, we are quite far from a complete theoretical understanding of the
problem. At the upper mass end ($M \ga 0.4M_\odot$) the inclusion of
molecular opacities in the model atmospheres, and the use of boundary
conditions from non-gray model atmosphere for the interior models has
improved considerably the match between the stellar models and the observed
colour of stars (see Baraffe et al. 1998, hereafter \cite{B98}). 
We have remarked that the optical
colours at \Teff$<$2500 K require grain inclusion in the models. This
is a difficult task, although insight is expected based on the ``dusty''
modeling by Allard (\cite{Allard}) 
and on the grain clouds approach by Burrows et al. (\cite{burrows98}).
Nevertheless, the unknowns in the modeling of grains (e.g. the
particle size distribution, the spatial distribution of clouds, their
composition, or the assumption of chemical equilibrium without rain--out) shift
very far away in time a self--consistent approach. In addition, the detailed
colour transition from L dwarfs to \CH4bds\ can not be yet described
quantitatively. 

Fig.~\ref{fighr} shows the evolutionary tracks of dwarfs and BDs, and a very
schematic separation between M, L and methane dwarfs, in terms of \Teff,
which we choose according to Reid et al. \cite{Reid}. These limits must be
taken with the caution due to the uncertainties described above.

\input{fighr}

\input{figobs}

\section{ Observational constraints on the density of faint \CH4bds }

In spite of all the difficulties in the models, the very blue J--K
colour, together with a very red I--K, should be a signature of the very low
\Teff, older BDs which we will call \CH4bds.
This simple recognition has prompted us to search for faint objects
characterized by small J--K but very large I--K or R--J, indicative of low
\Teff. A very useful data set comes from the ESO public survey data which are
targeted at regions far from the galactic plane and include the NTT SUSI deep
field ($l$=284$^o$, $b$=$+53^o$, Arnouts et al. \cite{NTTdeep1}, 
Moorwood et al.
\cite{NTTdeep2}), the EIS HDF--S field ($l$=328$^o$, $b$=$-49^o$, 
da Costa et al.
\cite{HDFS}) and the AXAF deep field
 ($l$=224$^o$, $b$=$-54^o$, Rengelink et al.
\cite{AXAF}). The overlap between infrared (K, J) and optical (I, R, V)
images cover a total area of 37 sq. arcmin and is complete to K=22.6,
J=24.4 and I=26.3 mag. These limits are somewhat unbalanced for our purposes,
i.e. too shallow in K and I for the depth reached in J. The colour selection
we use to extract methane dwarf candidates effectively translates into a cut
at J=23.1 and much above the limit of completeness of the J images. 

\input{tabcol}

The methane dwarf candidates were selected as those point--like objects with
J--K$<$0.5 and I--J$>$3, i.e. colours which are incompatible with any known
stellar source other than GL229B. This yielded three objects whose finding
charts and magnitudes are shown in Fig.~\ref{figobs}.
Considered the statistical errors,
the corresponding density ranges between $\approx$100 and $\approx$500
objects per sq. degree with magnitudes J$<$23.1.

In should be noted that the colours of the selected objects are also
compatible with quasars or QSOs at $z\!\sim\!9$, i.e. at redshifts where the
Ly$\alpha$ falls into the J band and the Lyman--break is moved beyond the I
photometric band. In such a case an observed magnitude J=23 would translate
into an absolute blue magnitude ${\rm M_B}\!\simeq\!-25$ and typical of
low--medium luminosity QSOs.
However, we consider this possibility unlikely because current estimates of
the density of quasars at various redshifts indicate a rapid drop beyond
$z$=3 with basically no object 
at $z\!>\!6$ (e.g. Shaver et al. \cite{shaver}).
Encouragingly, the \CH4bd\  nature of the brightest object was recently
confirmed by spectroscopic IR observations at the VLT (Cuby et
al.~\cite{cuby99}).\\

\section{Modelling of number counts and predicted colours}

As outlined in the introduction, it is not possible to model
self--consistently the whole BD evolution.
We decided therefore to adopt an admittedly rather simple--minded approach
to the colour derivation, in order get a first estimate of the expected number
counts of \CH4bds. On the other hand, we took care to model carefully the
galactic disk, so that the global model can be easily improved when more
precise evolutionary tracks will be available.

\input{figdist}

\subsection{The theoretical models}
As a first input, we need the time evolution of $\log L/L_\odot$ and \Teff\ 
for stellar masses which trace the low end of the main sequence and the
brown dwarf evolution.
We will use as a basis the models by Burrows et al. 
(\cite{burrows97} and private communication) 
employing non gray model atmospheres up to 1300K, and
similarly low resolution non gray models with grains up to 2400K. Above this
temperature, gray models with Rosseland opacities from Saumon are adopted.
These tracks are computed for masses from 0.237\Msun\ to 0.03\Msun, and are
evolved up to an age of 20Gyr, which is reached at quite low \Teff\ for the
smaller masses, so they constitute a very complete set to deal with all the
very low mass and brown dwarf regimes, from the M spectral type, through the
L dwarfs, to the postulated \CH4bds\ phase. Above this mass,
for the purpose of having predictions available also for somewhat hotter
stars, we adopt the tracks by D'Antona and Mazzitelli (\cite{DM97})
from 0.25 to 0.8\Msun.
Although the latter are computed with somewhat different inputs (e.g the
deuterium initial abundance is a half of the abundance assumed by Burrows et
al., the treatment of over--adiabatic convection is different) the merging of
the two sets is very reasonable, also because mainly the main sequence or
cooling part of the sequences are important to determine the stellar counts.
The gray atmosphere low mass models by D'Antona and Mazzitelli (\cite{DM97})
predict luminosities larger by $\sim 15$\% (generally $<$0.1mag) than the non
gray models e.g. by Chabrier \& Baraffe (1997, hereafter \cite{CB97}) and
have \Teff\ $\sim 2$\% larger. The Burrows et al models above 2400K compare
well with those of \cite{CB97} in luminosity, and they differ by  2\% in
\Teff.

\input{figcnt1}

\input{figcnt2}

\subsection{Colours and bolometric corrections}
\label{sect_col}
The magnitudes in I, J and K are not directly available in the track sets, so
we decided to adopt the admittedly crude procedure of estimating the I and J
band bolometric corrections ($BC$) 
from the \cite{B98} models, which are consistent
with the \cite{CB97} models. The \cite{B98} values have been used for \Teff
$>$ 2000~K. We
extrapolate these $BC$s to \Teff=600~K  using the models of
Burrows et al. (\cite{burrows97}). We
remark that the J band counts should be little affected by
uncertainties in the bolometric corrections. In fact, $BC_J$ is 2.19mag at
2000K, and very similar (2.23mag) at \Teff=600~K (see table~\ref{tabcol}). 
In other words, 25--30\%
of the flux for these objects emerges in the J band, regardless of their
temperature (e.g. Burrows et al. \cite{burrows98}). 

The colours J--K are taken from Burrows et al. \cite{burrows98} for 
\Teff$\leq$1000~K. Above 3300~K we adopt the colours of \cite{B98}. In the
intermediate range, these models fail reproduce the observations
of cooler dwarfs (Leggett et al. \cite{Leggett98}, \cite{B98}) and 
computations
including grains should be used (e.g. Allard \cite{Allard}, Jones \& Tsuji
\cite{jones}) to get the very red J--K colours down to GD165B. 
However, dusty models are still very incomplete and we therefore
employed a semiempirical approximation, imposing the models to pass through
the colour of GD165B and then to evolve to the blue.
Table~\ref{tabcol} shows the adopted relation between \Teff\  and
J--K. 
The reader should be well aware
that this approach represents an educated guess on the \CH4bds\ colour
evolution.
Nevertheless the semiempirical approach mainly influences the
{\it relative} number counts of the J versus K band, but not the \Teff\
selection, which however is itself is uncertain in the absence of
better modeling.

\subsection{The integrated number counts}
The number of stars within a given range of temperature (\Teff$\pm\!\Delta T$) 
and apparent magnitude (J$\pm\!\Delta$J)
expected from observations directed at a given direction $l,b$ (galactic
coordinates) and covering a relatively small solid angle $\Omega$ is given by
\begin{equation}
    \Sigma(\Teff,\J) = \int_0^\infty\! \Phi[\Teff,(\J\!-\!5\log R_{10})]
    \;\rho (R)\;\Omega\; R^2 dR
\end{equation}
where $R$ is the distance from the sun along the $l,b$ line of sight,
$R_{10}$=($R$/10$\;$pc) and $\rho$ is the total stellar density 
whose radial dependence can be approximated by
(Bachall \& Soneira, \cite{bahcall})
\begin{eqnarray*}
  \rho(R) & = & \rho_o\ \exp(-z/H-(r-r_o)/h)\\
   z \ \  & = & R\ \sin b \\
   r \ \  & = & \sqrt{(R\cos b)^2 - 2 R r_o \cos b \cos l + r_o^2}
\end{eqnarray*}
The coordinate $z$ is measured perpendicular to the galactic plane, $r$ is the 
radial distance from the galactic center
and $r_o$=8~kpc is the distance of the Sun.
The scale height is $H$=325~pc, consistent with the counts of M dwarfs,
and the scale length is $h$=3.5~kpc.
The local stellar density $\rho_o$ is given by
\begin{displaymath}
  \rho_o = 0.336 \int_{M_{min}}^{M_{max}} 
    \left({M\over0.1\, M_\odot}\right)^{-(1+x)}\ dM
\end{displaymath}
where we have adopted a power law initial mass function 
$dN/dM\propto M^{-(1+x)}$ and
the normalization factor at 0.1~\Msun\   is taken from
Reid \& Gizis (\cite{ReidGizis}). The mass limits considered here are
$M_{min}$=0.03~\Msun\  and $M_{max}$=0.8~\Msun.

In Eq.~(1) the term $\Phi(\Teff,{\rm M_J})$ gives the fraction of stars 
within the selected range of temperature and absolute J magnitude, i.e.
the relative density of stars on a given position of the theoretical
\Teff--${\rm M_J}$ HR diagram. This function, which
 depends on the shape of the IMF and on star formation history,
is constructed via a
Montecarlo technique assuming an uniform birth-rate from $t$=0 to the 
present disk age $t_{disk}$=10~Gyr.
The program randomly extracts masses (with probability given by the
chosen IMF shape) and ages from the track sets of Fig.~\ref{fighr}. Each
extraction provides a point in the $\rm M_{bol}$--\Teff\  
diagram which translates
into absolute magnitudes using the bolometric corrections of 
Table~\ref{tabcol}. 
The extracted stars
are grouped into suitably small bins of temperatures 
and $\rm M_J$, this yields the matrix which represents the 
function $\Phi$   which is used to compute the integral of Eq.~(1).

\input{figcol}

\input{tabcnt}

\subsection{Results}
The distribution of the apparent magnitude J as a function of the
distance (Fig.~\ref{figdist}) shows that the bulk of M dwarfs is at
magnitudes J$<$20, where BDs begin to appear. L--dwarfs are numerous at
J$\sim$20--22, while \CH4bds\ begin to be common only at J$>$21.

Fig.~\ref{figcnt1} and Fig.~\ref{figcnt2} also show the same feature.
The first displays the number counts per J magnitude 
within three mass bins: dwarfs
(0.5--0.7 \Msun), low mass dwarfs (0.08--0.14 \Msun) and brown dwarfs (below
0.08 \Msun). Being relatively
bright, the dwarfs show a maximum in the distribution at J$\simeq$17, the
lowest mass dwarfs also show a maximum at J$\simeq$20, while the BDs number
increases down to the limit at which we count them (J=25). %{\it
Therefore, either very deep searches are necessary to show an appreciable
number of \CH4bds, or a relatively deep survey covering most of the sky.
%}. 
Fig.~\ref{figcnt2} shows the total number counts versus J magnitude, 
putting into evidence which mass
ranges contribute to its different parts. The broad maximum in the counts at
J=17--19 includes only dwarfs and results from the 
combination of the effects due
to the IMF, the mass luminosity relation for the main sequence and the
radial density distribution. The rising of the number counts at J$>$22 is
instead due to the low mass brown dwarfs. Finally, Fig.~\ref{figcol} shows
simulated J vs J--K colour--magnitude diagram for observations covering 1 sq.
degree of sky. The large scatter of J--K colours for the \CH4bds\ reflects
the effect of methane bands absorption on the J--K colour, which is expected
to increase rapidly between \Teff=1000~K (J--K=$-0.1$) and \Teff=600~K
(J--K=$-1.4$, see table~\ref{tabcol}).

The resulting counts are tabulated in Table~\ref{tabcnt} for galactic
coordinates similar to those of the ESO fields, namely $b$=$55^o$
$l=$340$^o$, and at the galactic pole ($b$=90$^o$).

\subsection{Comparison with observations}
Table~\ref{tabcnt} shows that a considerable number of L--dwarfs are predicted,
but practically no \CH4bds (at most 1 in 1000 sq. degrees), in a large area
limited at K$<$14.5, in agreement with the early results of 2MASS. The
very recent observations of four relatively bright \CH4bds\ (Strauss et
al.~\cite{strauss99}, Burgasser et al.~\cite{burgasser99}) yields a surface
density of $\simeq$0.002 cool methane dwarfs with J$<$16 and a factor of
3--6 lower than those predicted here using flat and steep IMFs, respectively,
and considering only the density of objects of \Teff$<$1400 K. However, these
differences are not necessarily significant, given the small number of
objects detected and the possibility that the the above searches may not be
yet complete.

Our observed density of 3 objects with J$<$23.1 within 37 sq. arcmin is
consistent (i.e. within the Poisson's noise) with the space density predicted
using a relatively steep IMF ($x=1$) while it is a factor of 4$\pm$2 larger
than expected from a flatter IMF ($x=0$). Remember that this latter IMF is
certainly more consistent with the slope derived from nearby low mass stars
(Reid and Gizis 1997). Given the very small number of objects detected we
consider it premature to draw strong conclusions about the IMF.

\section{Conclusions}

We presented three \CH4bds\ candidates with J$<$23.1 found in an area of 37
sq. arcmin covered by the ESO public images survey and modelled the expected
number counts for these objects as well as for L and M dwarfs.
Theoretical predictions are made on the basis of a sound galactic model
including the most complete set of today's available evolutionary tracks and
colours--\Teff\  correlations.  The model could be significantly improved only
when self--consistent computations will be available, including model
atmospheres tested on the methane dwarfs which will be discovered.

The predicted surface densities are in reasonably good agreement with the
observed counts of L dwarfs and with the sparse observational data presently
available for methane dwarfs, namely the lack of objects with K$<$14.5 in
the early 2MASS survey, the recent finding of four relatively bright
(J$<$16) \CH4bds\ in 1800$^2$ sq. degrees  and the three much fainter
candidates discussed here.

We also showed that, due to their very low intrinsic luminosities, methane
dwarfs become numerous (relative to L and M dwarfs) only at very faint J
magnitudes (see figures from ~\ref{figdist} to ~\ref{figcol}). To significantly
improve the observational database it is therefore necessary to cover a
reasonably large area of sky (several sq. degrees) down to J$\simeq$22, as
well as K$\simeq$22 and I$\ga$27, the latter two being necessary for
selecting the \CH4bd\ candidates. We look forward for this type of
observations to be soon performed by the IR large field deep imagers
operating or planned on large telescopes.

\begin{acknowledgements}
We are grateful to ESO for making the EIS and NTT SUSI Deep field data
available to the community. The EIS observations have been carried out using
the ESO New Technology Telescope (NTT) at the La Silla observatory under
Program-ID Nos. 59.A-9005(A) and 60.A-9005(A). This work was partly supported
by the Italian Ministry for University and Research (MURST) under grant
Cofin98-02-32. We thank A. Burrows for helpful discussion and G. Chabrier
for a careful referee's report.
\end{acknowledgements}

\end{document}

%% file: fighr.tex
\begin{figure}
\centerline{\resizebox{8.8cm}{!}{\rotatebox{0}{\includegraphics{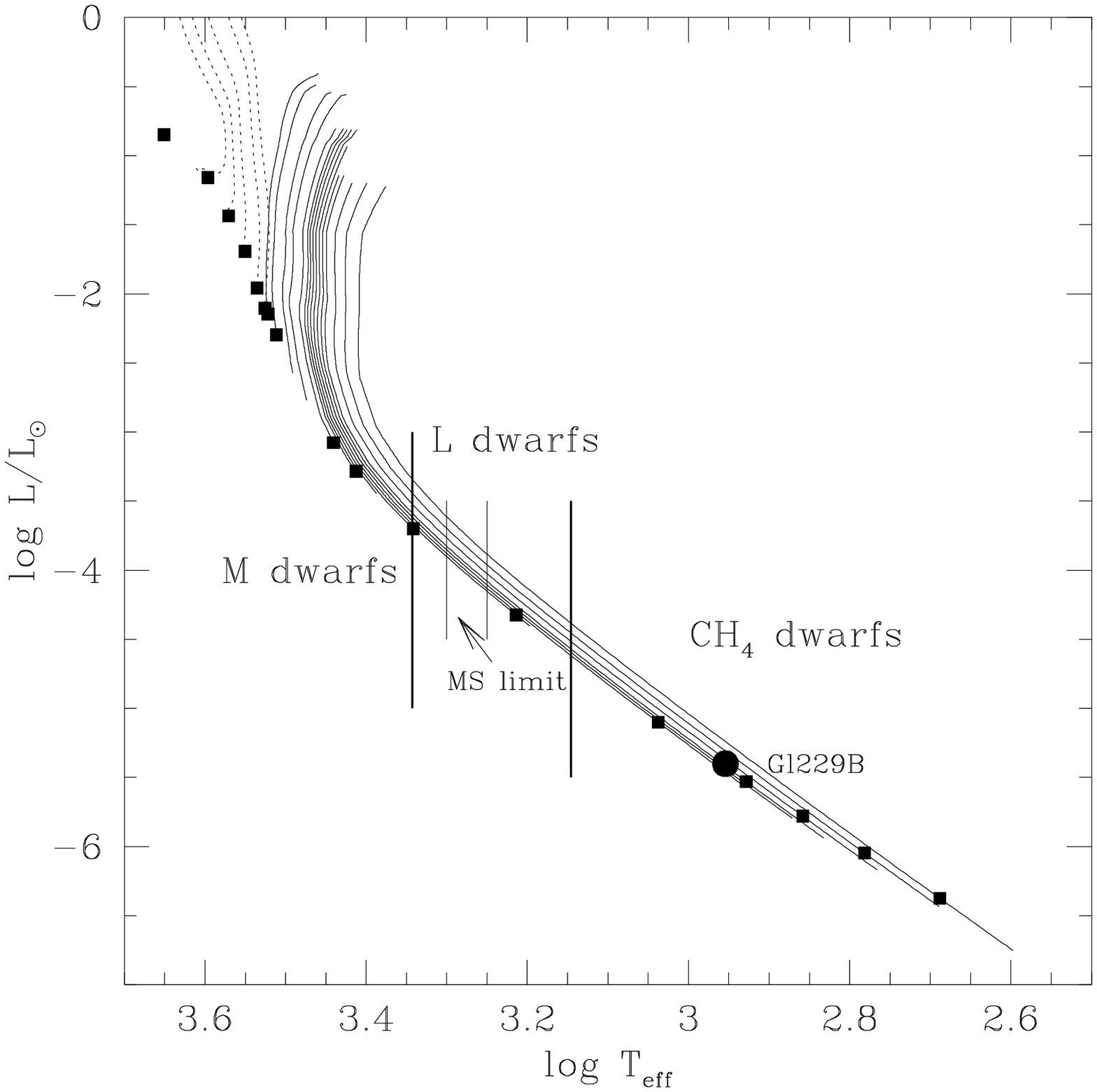}}}}
\caption[]{
HR diagram with the evolutionary tracks of low mass stars.
The dashed lines are from
 D'Antona \& Mazzitelli (\cite{DM97}) and cover the 0.6 to 0.2~\Msun\
range, while the tracks for lower mass stars 
(0.237 to 0.03~\Msun) are from Burrows et al. (\cite{burrows98}).
The square symbols mark the positions at 10$^{10}$ years.
The approximate boundaries of the M, L and methane dwarfs are
shown. The Main Sequence limit is between 2000 and 1800K depending on the
opacities adopted, so it falls in the middle of the L spectral type.
\label{fighr}
}
\end{figure}

%% file: figobs.tex
\begin{figure}
\resizebox{8.8cm}{!}{{\rotatebox{0}{\includegraphics{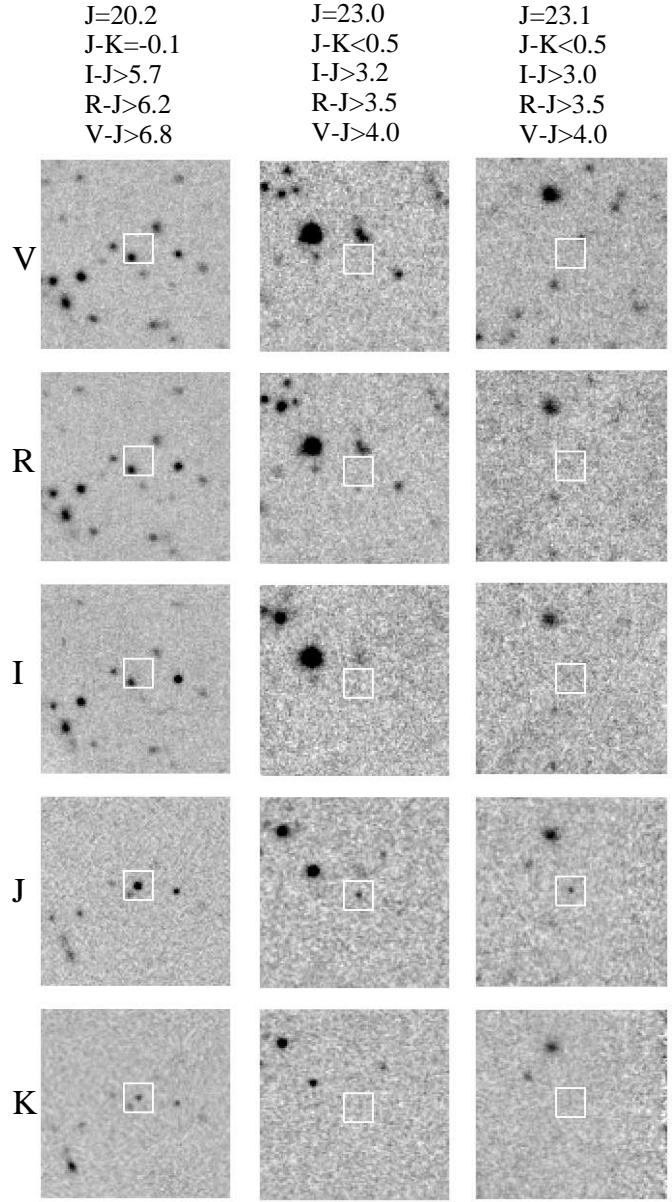}}}}
\caption
{Selected regions of the ESO deep fields showing the \CH4bds\
candidates. Each sub-frame is 28"x28" with N to the left and E to the
bottom, the position of the candidate is marked by a white rectangle.
The object magnitudes and colours are $\pm$0.15 while the upper limits
are at the 3$\sigma$ level.
 \label{figobs}
}
\end{figure}

%% file: tabcol.tex
\begin{table}
%  \caption[]{Adopted conversions \Teff\  versus J--K and $BC_J$ }
  \caption[]{Adopted colours and bolometric corrections}
%   \begin{tabular*}{7.cm}{cccc}   \hline
   \begin{tabular}{cccc}   \hline
 & & & \\
~~~~$\log \Teff$ & $BC_J^{(1)}$  & 
   (J--K)$_{mod}^{(2)}$ & (J--K)$_{emp}^{(3)}$ \\
 & & & \\
	   \hline
 2.768    &  2.23\rlap{$^a$}  &  \llap{--}1.44\rlap{$^a$} & \llap{--}1.44  \\
 3.000    &  2.20\rlap{$^a$}  &  \llap{--}0.10\rlap{$^a$} & \llap{--}0.10  \\
 3.301    &  2.19  &    0.87     &  1.37  \\
 3.362    &  2.16  &    0.92     &  1.21  \\
 3.423    &  2.07  &    0.93     &  1.06  \\
 3.447    &  2.00  &    0.92     &  0.97  \\
 3.498    &  1.88  &    0.87     &  0.91  \\
 3.529    &  1.81  &    0.84     &  0.84   \\
 3.554    &  1.75  &    0.81     &  0.81   \\
 3.592    &  1.66  &    0.79     &  0.79   \\
 3.631    &  1.54  &    0.71     &  0.71   \\
 3.674    &  1.39  &    0.57     &  0.57   \\
 3.713    &  1.26  &    0.43     &  0.43   \\
	   \hline
%   \end{tabular*}
   \end{tabular}
\begin{list}{}{}
\item[$^{(1)}$] Bolometric correction in J, from the models of Burrows et al.
(\cite{burrows98}) except for those in note $a$
\item[$^{(2)}$] Colours predicted by models, from \cite{B98} except for those
in note $a$
\item[$^{(3)}$] Empirical colours adopted here, see Sect.~\ref{sect_col}
\item[$^a$] Values from Burrows et al. (\cite{burrows97})
\end{list}
\label{tabcol}
 \end{table}

%% file: figdist.tex
\begin{figure}
\centerline{\resizebox{8.8cm}{!}{\rotatebox{0}{\includegraphics{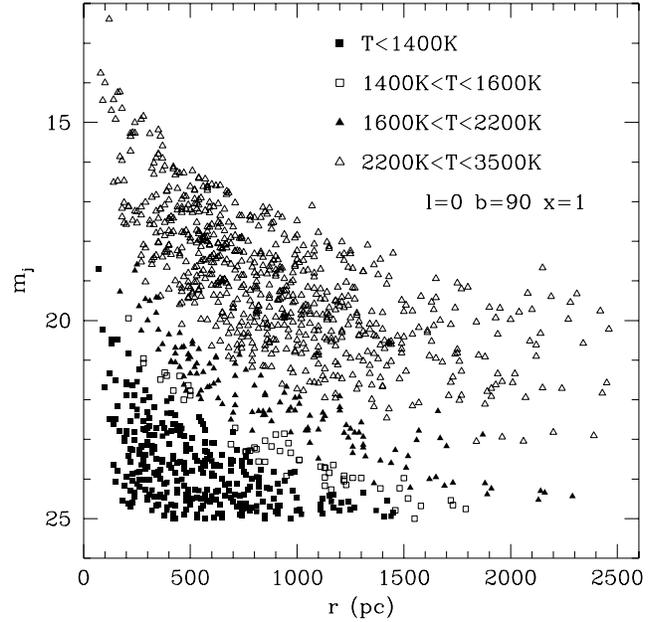}}}}
\caption {Example of the predicted distribution of apparent stellar J
magnitudes versus the distance, for simulated observations covering 1 square
degree.
 } \label{figdist} \end{figure}

%% file: figcnt1.tex
\begin{figure}
\centerline{\resizebox{8.8cm}{!}{\rotatebox{0}{\includegraphics{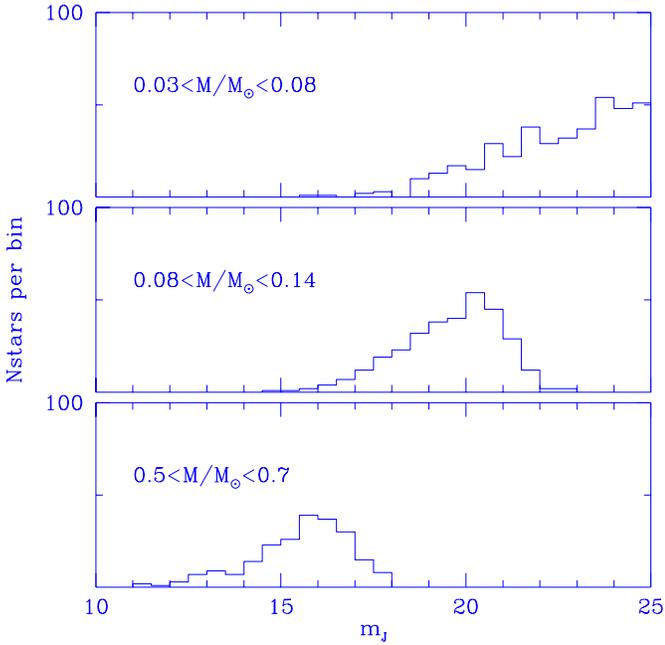}}}}
\caption[]{ Stellar counts (in bins of 0.5mag, per square degree) as function
of the apparent J magnitude for three selected mass ranges.} 
\label{figcnt1}
\end{figure}

%% file: figcnt2.tex
\begin{figure}
\centerline{\resizebox{8.8cm}{!}{\rotatebox{0}{\includegraphics{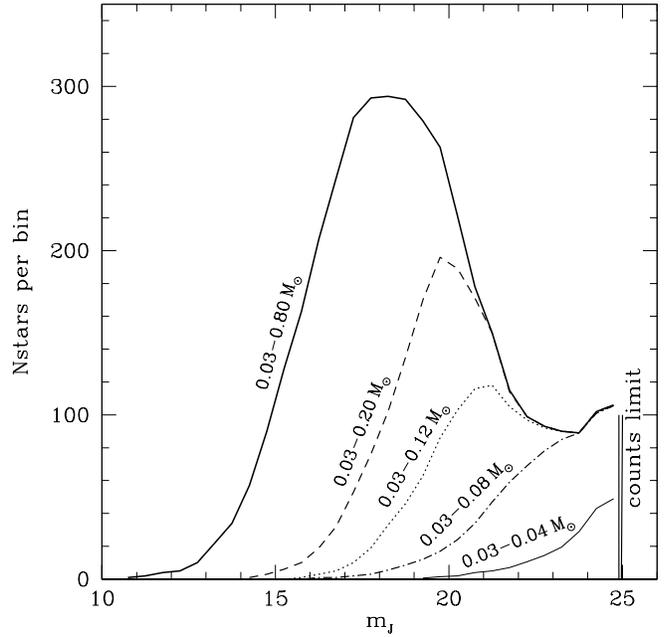}}}}
\caption {Stellar counts in bins of half magnitude and per square degree
as a function of the apparent magnitude J, in the case l=0, b=90, x=0. The
cumulative counts from 0.03 to 0.8\Msun are
shown in the thick line. Partial counts due to separate mass ranges
are also given to  show at which magnitudes each mass range contributes
mostly to the counts. The continuous line at the lower right represents the
counts from
0.03 to 0.04\Msun, the dash-dotted one from 0.03 to 0.08\Msun, the
dotted line from 0.03 to 0.12\Msun and the dashed from 0.03 to 0.2\Msun
The BD mass range is largely represented only
for J$>$20. The counts are limited to J=25, but BDs would be present
also above this magnitude. 
\label{figcnt2} } \end{figure}

%% file: figcol.tex
\begin{figure}
\centerline{\resizebox{8.8cm}{!}{\rotatebox{0}{\includegraphics{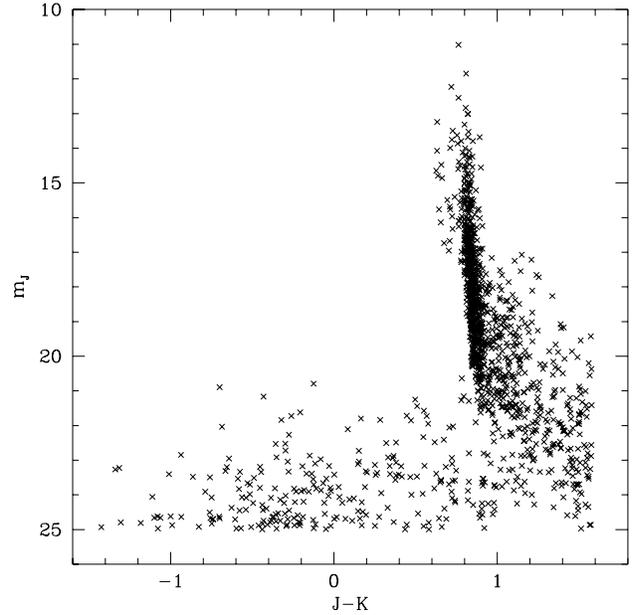}}}}
\caption {Example of the predicted distribution of apparent stellar J
magnitudes versus J--K colour for observations covering of 1 square degree.
The points cluster along the main sequence defined by the thick part along
J--K$\simeq$1, then we recognize the L--dwarfs which populate the region at
1$<$J--K$<$1.6, before evolving to much bluer colours when the methane bands
begin to appear in the atmosphere. } 
\label{figcol} \end{figure}

%% file: tabcnt.tex
\begin{table*}
  \caption[]{Predicted brown dwarf densities$^{(1)}$}
%   \begin{tabular*}{15.3cm}{|c|ccccc|ccccc|}   \hline
   \begin{tabular}{|c|ccccc|ccccc|}   \hline
%	\medskip
      J$_{lim}$ &  \CH4bds$^a$  & ~L~$^b$  & M$^c$  & M$^d$ & M$^e$ &
 \CH4bds$^a$ & ~L~$^b$  & M$^c$ & M$^d$ & M$^e$  \\
	   \hline
	   \hline
IMF$^f$ $x$=0.0 & \multicolumn{2}{c}{$l$=0$^o$~~~$b$=90$^o$} &&&&
\multicolumn{2}{c}{$l$=340$^o$~~~ $b$=55$^o$} & & & \\            \hline
14.0 & .0004 + .0003   & .004   & .01 &  .3  &  7    
     & .0004 + .0003   & .004   & .01 &  .3  &  7   \\

15.0 & .001  + .001    & .016   & .05 &  1   & 20    
     & .001  + .001    & .016   & .05 &  1   & 25   \\

16.0 & .006  + .005    & .06    & .19 &  4   & 65    
     & .006  + .005    & .06    & .19 &  5   & 78   \\

17.0 & .022  + .019    & .20    & .65 &  12  & 160   
     & .023  + .020    & .21    & .80 &  13  & 215  \\

18.0 & .085  + .074    & .78    & 2.1 &  33  & 320
     & .088  + .079    & .84    & 2.4 &  40  & 490  \\

19.0 & .32   + .27     & 2.6    & 6.4 &  79  & 515
     & .33   + .30     & 2.9    & 7.6 &  106 & 900  \\

20.0 &  1.2 + .91      & 7.7    & 16  &  150  & 650
     &  1.3 + 1.0      & 9.1    & 22  &  240  & 1300 \\

21.0 &  4.0 + 2.7      & 19     & 34  &  230  & 702
     &  4.4 + 3.2      & 25     & 51  &  420  & 1500 \\

22.0 &   13 + 7.5      & 40     & 54  &  280  & 710
     &   15 + 9.3      & 60     & 95  &  570  & 1550 \\

23.0 &   38 + 16       & 74     &  73 &  290  & 710
     &   50 + 20       & 115    & 130 &  570  & 1550 \\

24.0 &   90 + 30       & 90     &  74 &  290  & 710
     &   112 + 40      & 165    & 145 &  575   & 1550 \\

25.0 &   160 + 38      & 90     &  74   & 290 & 710
     &   270 + 72      & 200    & 150   & 575 & 1550  \\
	   \hline
	   \hline
IMF$^f$ $x$=1.0 & \multicolumn{2}{c}{$l$=0$^o$~~~$b$=90$^o$} &&&&
\multicolumn{2}{c}{$l$=340$^o$~~~ $b$=55$^o$} & & & \\            \hline
	   \hline
14.0 &  .0008 + .0006 & .006 & .018 &  .3   & 3 
     &  .0008 + .0006 & .006 & .018 &  .3   & 3  \\

15.0 &  .003  + .002  & .025 & .068 &  1    & 9 
     &  .003  + .002  & .025 & .070 &  1    & 9   \\

16.0 & .01  + .009   & .1   & .25  &  4    &  32 
     & .01  + .009   & .1   & .25  &  4    &  32  \\

17.0 & .045 + .036   & .31  & .9   &  12   &  68 
     & .046 + .036   & .32  & .9   &  13   &  90  \\

18.0 &  .17  + .13   & 1.2  &  3   &  32   & 140
     &  .18  + .14   & 1.3  &  3   &  40   & 210 \\

19.0 &  .69  + .48   & 4.0  &  8.5 &  75   & 240 
     &  .72  + .54   & 4.5  &  10  &  105  & 400 \\

20.0 &  2.3 + 1.6    & 12   & 22   &  150  & 310  
     &  2.5 + 1.8    & 14   & 28   &  230  & 600 \\

21.0 &  8.0 + 4.9    & 30   & 44   &  225  & 345 
     &  9.0 + 5.7    & 39   & 66   &  400  & 720 \\

22.0 &  25 + 13      & 61   & 69   & 270   & 350 
     &  29 + 17      & 94   & 120  & 550   & 760 \\

23.0 &  84 + 24      &  99  & 73   & 270   & 350 
     &  95 + 47      & 170  & 150  & 550   & 760 \\

24.0 & 200 + 38      & 125  & 76   & 270   & 350 
     & 240 + 95      & 235  & 170  & 560   & 760 \\

25.0 & 385 + 52      & 130  & 76   & 270   & 350 
     & 535 + 130     & 270  & 170  & 560   & 760\\
	   \hline
%   \end{tabular*}
   \end{tabular}
\begin{list}{}{}
\item[$^{(1)}$] Cumulative number of objects per square degree with
magnitude  J$\le$J$_{lim}$
\item[$^{\mathrm{a}}$] \Teff$<$1400~K + 1400~K$<$\Teff$<$1600~K
\item[$^{\mathrm{b}}$] 1600~K$<$\Teff$<$2200~K
\item[$^{\mathrm{c}}$] 2200~K$<$\Teff$<$2500~K
\item[$^{\mathrm{d}}$] 2500~K$<$\Teff$<$3000~K
\item[$^{\mathrm{e}}$] 3000~K$<$\Teff$<$3500~K
\item[$^{\mathrm{f}}$] IMF in the form $dN/dM \propto M^{-(1+x)}$
\end{list}
\label{tabcnt}
 \end{table*}